\documentclass[twocolumn]{aastex6}

\usepackage{graphicx}
\usepackage{amssymb}
\usepackage{amsmath}

\usepackage{natbib}
\usepackage{txfonts}
\usepackage{multirow}
\usepackage{array}
\usepackage{sidecap}
\usepackage{hyperref}
\usepackage{epstopdf}
\usepackage{footnote}
\usepackage{tabularx}
\usepackage{booktabs}
\usepackage{longtable}
\usepackage{tabu}
\newcommand{\Hinode}{\textit{Hinode}}

\newcommand{\sdo}{\textit{SDO}}

\shortauthors{Panesar et al.}

\begin{document}
	\title{\bf Magnetic Flux Cancelation as the Origin of Solar Quiet Region Pre-Jet Minifilaments}
	%Magnetic Flux Cancellation is the Buildup of Solar Pre-Jet Minifilament-Holding Magnetic Arcades

	\author{Navdeep K. Panesar\altaffilmark{1}, Alphonse C. Sterling\altaffilmark{1}, Ronald L. Moore\altaffilmark{1,2}}
	\affil{{$^1$}Heliophysics and Planetary Science Office, ZP13, Marshall Space Flight Center, Huntsville, AL 35812, USA}
	\affil{{$^2$}Center for Space Plasma and Aeronomic Research (CSPAR), UAH, Huntsville, AL 35805, USA}
	\email{navdeep.k.panesar@nasa.gov}
	
	%\and

	\begin{abstract}
		We investigate the origin of ten solar quiet region pre-jet \textit{minifilaments}, using EUV images from \textit{SDO}/AIA and  magnetograms from \textit{SDO}/HMI. We recently found \citep{panesar16b} that quiet region coronal jets are driven by minifilament eruptions, where those eruptions result from flux cancelation at the magnetic neutral line under the minifilament. Here, we study the longer-term origin of the pre-jet minifilaments themselves. We find that they result from flux cancelation between minority-polarity and majority-polarity flux patches. In each of ten pre-jet regions, we find that  opposite-polarity patches of magnetic flux converge and cancel, with a flux reduction of 10--40\% from before to after the minifilament appears. For our ten events, the minifilaments exist for periods ranging from 1.5 hr to two days before erupting to make a jet. Apparently, the flux cancelation builds highly sheared field that runs above and traces the neutral line, and the cool-transition-region-plasma minifilament forms in this field and is suspended in it. We infer that the convergence of the opposite-polarity patches results in reconnection in the low corona that  builds a magnetic arcade enveloping the minifilament in its core, and  that the continuing flux cancelation at the neutral line finally destabilizes the minifilament field so that it erupts and drives the production of a coronal jet. Thus our observations strongly support that quiet region magnetic flux cancelation results in both the formation of the pre-jet minifilament and its jet-driving eruption.

	\end{abstract}

\section{INTRODUCTION}

Eruptive small-scale features, such as \textit{minifilaments}, are frequently observed on the solar surface \citep{moore77,hermans86,chae99,wang00,sakajiri04,zuccarello07,ren08}. They are located above a  magnetic neutral line (also known as polarity inversion line, PIL) between adjacent opposite-polarity magnetic flux clumps in the photosphere \citep{martin86,mar98}. On average, minifilaments have projected lengths of about 20 $\times$ 10$^{3}$ km \citep{wang00,sterling15,panesar16b} and are analogous to large-scale solar filaments, which  reside in the highly sheared magnetic fields in and above filament channels \citep{mart66,gai97,mar98}.

Coronal jets are transient collimated features \citep{raouafi16} typically observed in extreme ultraviolet (EUV) \citep{ywang98} and X-ray \citep{shibata92,canfield96,alexander99} emission. They occur in different solar environments over a broad range of temporal and spatial scales \citep{cirtain07,moore10,innes16}. Coronal jets typically show a brightening at an edge of their base (called the jet-base bright point, JBP) and a bright spire. Jets are typically observed at sites of emerging and canceling magnetic flux  \citep{shen12,panesar16a,sterling16,chandra17}. The jet spire is produced by magnetic reconnection of closed field in the base with ambient open or far-reaching field; the spire plasma is driven out along the reconnected open/far-reaching field. Several researchers  \citep[e.g.][]{hong11,shen12,sterling15,sterling16,panesar16b} have presented evidence that coronal jets are driven by minifilament eruptions. Some studies found that these minifilaments erupt from a neutral line at which unambiguous magnetic flux cancelation is occurring \citep{hong11,huang12,adams14,young14b,young14a}.

%%%%%%%%%%%%%%%%%%%%%%%%%%%%%%%%%%%%%%%%%%%%%%%%%%%%%%%%%%%%%%%%%%%%%%

\floattable
\begin{table*}
	\begin{center}
		\setlength{\tabcolsep}{.70em}
		%	\footnotesize
		\caption{Measured parameters for the observed quiet region pre-jet minifilaments \label{tab:list}}
		%	\small
		\renewcommand{\arraystretch}{1.0}% for Tighter rows
		\begin{tabular}{c*{9}{c}}
			\noalign{\smallskip}\tableline\tableline \noalign{\smallskip}
			%\multicolumn{10}{l}
			Event &  Minifil. formation\tablenotemark{a} &  Minifil. eruption\tablenotemark{b} &Location\tablenotemark{c}  & Duration of \tablenotemark{c}&  Width of\tablenotemark{d}  & No. of\tablenotemark{e}  & $\Phi$ values\tablenotemark{f} &\% of $\Phi$\tablenotemark{g}\\
			
			No.   &    time (UT)   & time (UT)   &helio. cord. & minifil. (hrs) &  minifil. (km) & Jets & 10$^{19}$ Mx & reduction  \\
			
			\noalign{\smallskip}\hline \noalign{\smallskip}
			J1 & 2012 Mar 21   22:46  & 2012 Mar 22~ 04:46 & S09, E29  &  6.0 &  2000$\pm$500 &  1 & 1.6 & 20 $\pm$ 6.8 \\   
			
			J2 & 2012 Jul 01   05:58  &  2012 Jul 01 08:29    & N12, E02 & 2.5 &  1500$\pm$200  &  1 &  1.9\tablenotemark{h} &  20 $\pm$ 7.3      \\   
			
			J3 & 2012 Jul 07   ---\tablenotemark{i}  & 2012 Jul 07 21:31 & S15, E12  & -- &  2200$\pm$200 &  1 &  -- & --  \\   [-0.5ex]
			
			J4 & 2012 Aug 04 05:14  & \textit{2012 Aug 05 01:58}\tablenotemark{j}, & N07, E30  & 21 &  2500$\pm$500 & 2  &  5.8 & 14 $\pm$ 4.6  \\   [-0.5ex]
			&  & 2012 Aug 05  02:20  &  &  &  &  \\   
			J5 & 2012 Aug 10  19:43  & 2012 Aug 10  23:03 & S31, E11   & 3.2 &  1500$\pm$200 &  1  &  0.9 &  27 $\pm$ 6.1  \\ 
			
			J6 & 2012 Sept 19   17:15  & 2012 Sept 20  22:52    & S34, E11  & 34 &  2500$\pm$500 & 2 &  3.0 &  9 $\pm$ 5.3      \\  
			
			J7 & 2012 Sept 21  00:51  & 2012 Sept 21 03:33  & S34, E08 & 3.5  &  2500$\pm$500  & 1 &  1.7 &  38 $\pm$ 2.6  \\   
			
			J8 & 2012 Sept 21   23:55  & 2012 Sept 22 01:25 & N01, E20  &   1.5 &  1500$\pm$500 & 1&  0.9 & 38 $\pm$ 5.5  \\ 
			
			J9 & 2012 Nov 11 02:56  & \textit{2012 Nov 11 13:08},  & S23, E01 &   49.5 &  2500$\pm$500 & 4  &  --\tablenotemark{k} & --  \\ 
			&  & \textit{2012 Nov 12 17:06},  &  &  &  &  \\  
			&  & \textit{2012 Nov 12  21:34},  &  &  &  &  \\
			&  & 2012 Nov 13  04:20  &  &  &  &  \\
			J10 & 2012 Dec 13   08:06  & \textit{2012 Dec 13 10:11},   & S01, W01 &   2.5 &  1600$\pm$200 & 2 &  1.2 & 7.0 $\pm$ 8.3 \\ 
			&  &2012 Dec 13 10:36  &  &  &  &  \\ 
			\noalign{\smallskip}\tableline\tableline \noalign{\smallskip}
			
		\end{tabular}
		
		\tablenotetext{a}{Approximate time of minifilament formation based on the appearance of dark cool material in AIA 304 and 171 \AA\ images.}
		\tablenotetext{b}{Approximate time of the minifilament eruption.} 
		\tablenotetext{c}{Approximate location of the jet region on the solar disk in heliographic coordinates.} 
		\tablenotetext{c}{Total duration or lifetime of the minifilament visibility in AIA 304 and 171 \AA\ images. The duration is almost the same in these wavelengths.}
		\tablenotetext{d}{Measured within 30--60 minutes before the final jet eruption.}
		\tablenotetext{e}{Total number of jets from the same PIL where minifilaments were located before eruption. }
		\tablenotetext{f}{ Average flux ($\Phi$) values of the minority flux clumps 2-3 hours before minifilament formation. }
		\tablenotetext{g}{Flux  change between 2-3 hours before minifilament formation and 0-1 hours after formation. }
		\tablenotetext{h}{Here we measured positive flux. In \cite{panesar16b}, for the same event we measured the negative flux. }
		\tablenotetext{i}{Ambiguous.}
		\tablenotetext{j}{Minifilament starts to move before this time.}
		\tablenotetext{k}{Unable to estimate the flux cancelation values because of the concurrent flux emergence at a neighboring location.  }
		
	\end{center}
\end{table*}

%%%%%%%%%%%%%%%%%%%%%%%%%%%%%%%%%%%%%%%%%%%%%%%%%%%%%%%%%%%%%%%%%%%%%%%%%%%%%%%%%%%%%%%%
\cite{panesar16b} investigated the triggering of ten on-disk quiet region jets, and found that flux cancelation led to a jet-driving minifilament eruption in all ten events. The magnetic flux cancelation occurred at the neutral line under the minifilament between  majority-polarity and minority-polarity magnetic flux patches. They inferred that the cancelation finally destabilizes the field holding the minifilament, and that the eruption then ensues as proposed in \cite{sterling15}: (1) The minifilament starts to rise. As it rises, \textit{ internal reconnection} occurs between opposite-polarity legs of the field enveloping the minifilament, producing the JBP, a miniature flare arcade that straddles the neutral line of the pre-eruption minifilament; that is, the JBP is a miniature version of the flare arcade that is made in large-scale filament eruptions. (2) The jet spire is made by the erupting-arcade field enveloping the minifilament barging into ambient oppositely-directed far-reaching field and undergoing  \textit{external reconnection} with it. (3) The external reconnection opens the erupting closed minifilament field, and the minifilament's cool-transition-region plasma is driven out along the newly-opened field lines and appears as part of the jet spire. Except for the spire-producing external reconnection, jets and their accompanying JBP are driven by minifilament eruptions in a manner analogous to how larger filament eruptions drive the production of a solar flare and  coronal mass ejection (CME).

Important outstanding questions regarding jets include: How and when are pre-jet minifilaments formed? What is the evolution of the magnetic field that leads to their formation? Do they appear just before the jet eruption, or a day or so before? 

More broadly, because the evidence strongly indicates that minifilaments are miniature counterparts of large-scale filaments \citep[e.g.][]{hermans86,wang00,sterling15}, by studying the formation of minifilaments we expect that we can also learn about the formation mechanism of large-scale filaments.

In this paper, we investigate the formation mechanism of ten on-disk pre-jet minifilaments in quiet regions. These are the minifilaments in the ten jets that we analyzed in \cite{panesar16b}; in that paper we found that the  jet-producing eruptions of these minifilaments were triggered by flux cancelation, while here we study the formation and pre-jet evolution of the minifilaments. We again use EUV images from the \textit{Solar Dynamics Observatory (SDO)}/Atmospheric Imaging Assembly (AIA) and photospheric magnetograms from the \sdo/Helioseismic and Magnetic Imager (HMI). We mainly investigate the magnetic field evolution by tracking the changes in magnetic flux that lead to the formation of the minifilament. We find that all the pre-jet minifilaments in our study arise from flux cancelation at the neutral line on which the minifilament forms.

\begin{figure*}
	\centering
	\includegraphics[width=\linewidth]{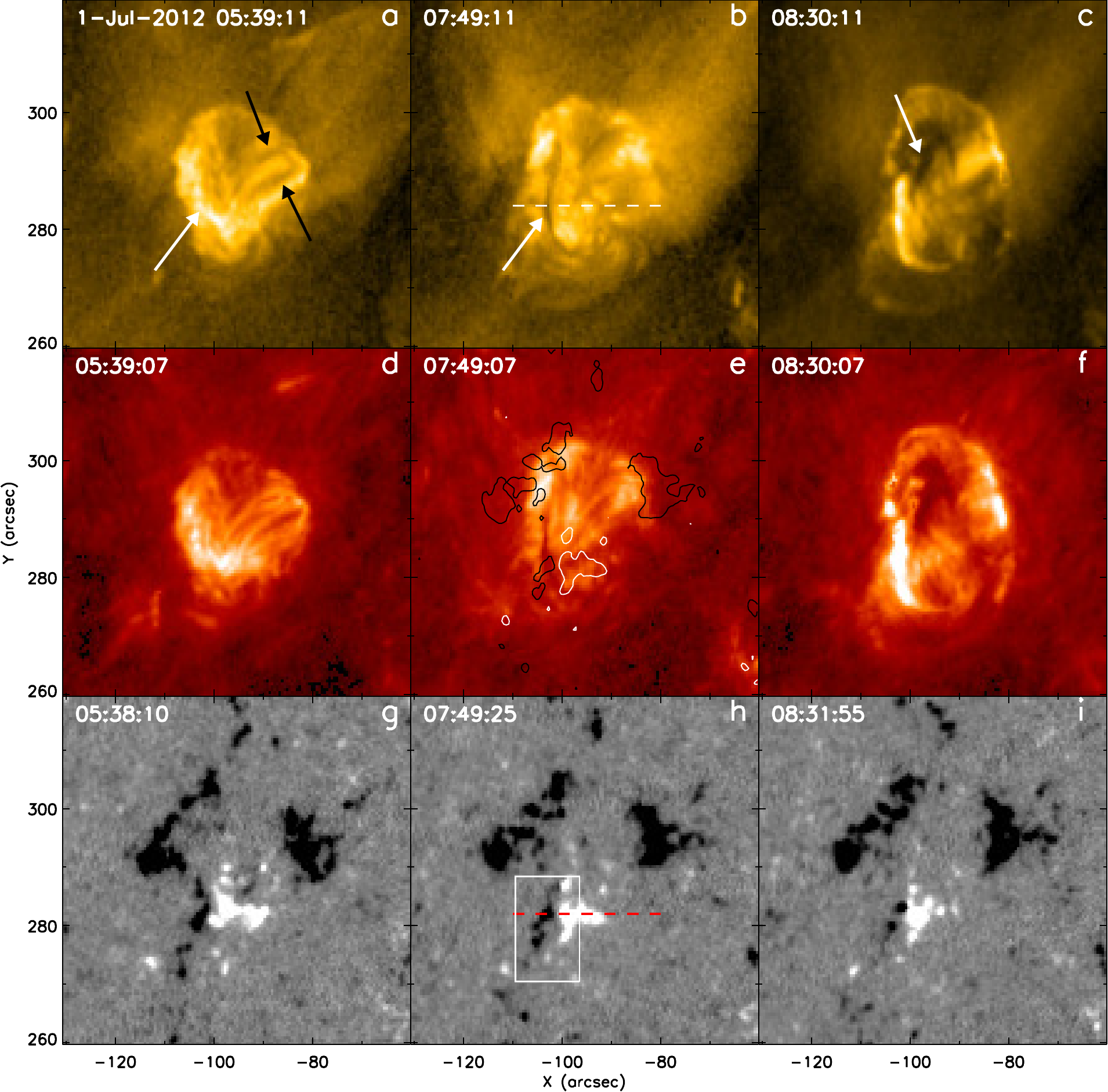}
	\caption{Region of the quiet-region minifilament that led to jet J2 of Table \ref{tab:list}, showing (a-c) 171 \AA\ and (d-f)
		304 \AA\ intensity images from AIA, and (g-i) magnetograms of the same region from HMI\@.  Dates at the top of each column apply to each of the three panels in those respective columns.  Left, middle, and right columns respectively show the region before minifilament formation, a time when the minifilament is present, and a time when the minifilament is erupting to form the jet. In (a), the black arrows point to  dark features (field transition arches), and the white arrow shows brightenings at the neutral line before the minifilament formation. In (b), the arrow shows the minifilament and the white dashed line shows the east-west cut for the time-distance map of Figure \ref{fig1b}a. The arrow in (c) points to the minifilament during eruption onset. The  boxed region in (h) shows the area used for measuring the negative magnetic flux for the plot shown in Figure \ref{fig1b}b; the red dashed line shows the east-west cut for the Figure \ref{fig1b}c time-distance map. HMI contours (level $\pm$ 50 G) of (h)  are overlaid onto panel (e), where white and black respectively represent positive and negative
		polarities. North is in the top and west is in the right of all the images and movies.
		Animations (MOVIE1) of this Figure are available.} \label{fig1a}
\end{figure*}

\begin{figure}
	\centering
	\includegraphics[width=\linewidth]{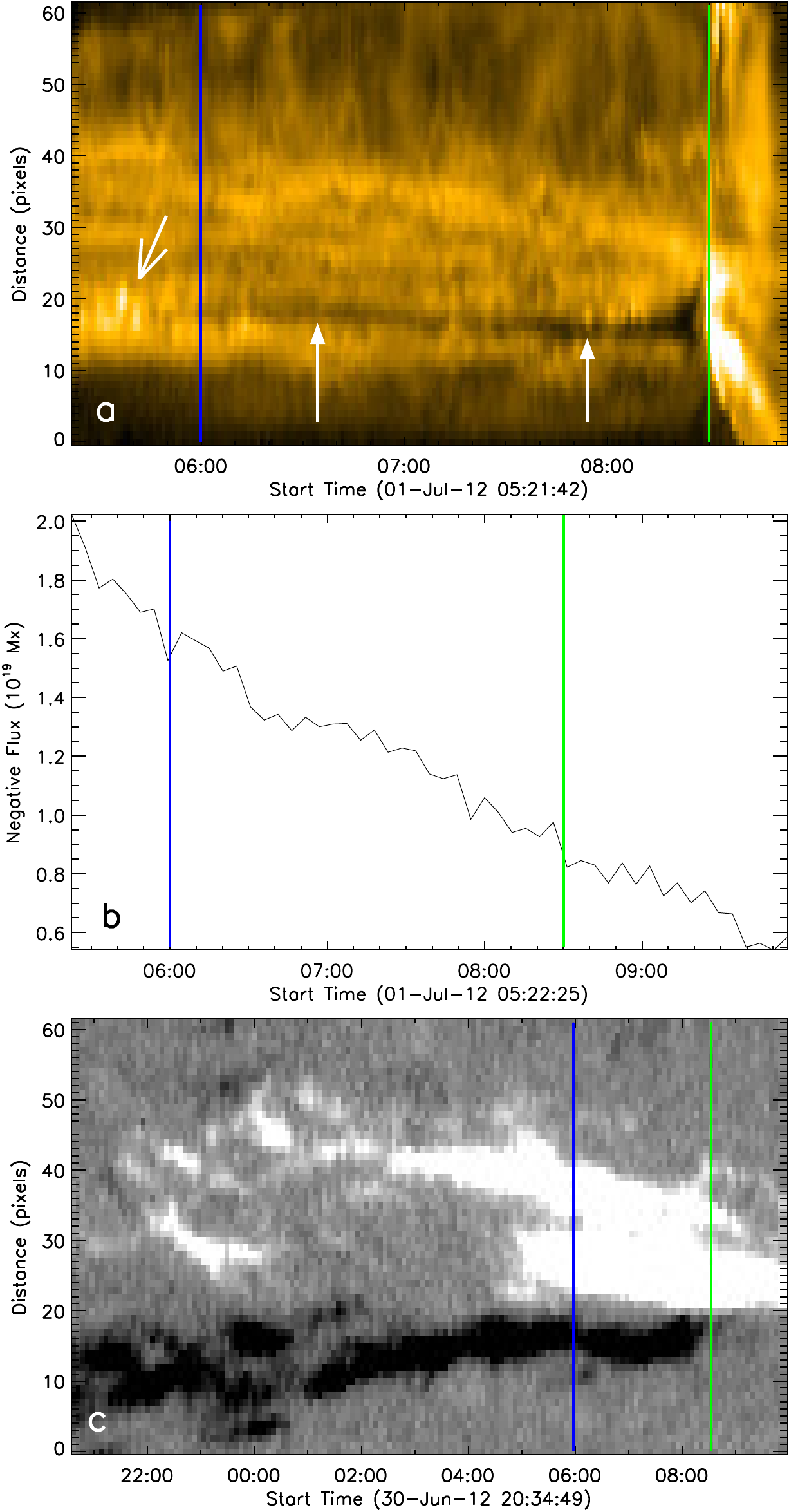}
	\caption{Minifilament formation and flux cancelation for jet J2: (a) 171 \AA\  time-distance map of the pixels along the white dashed line of Figure \ref{fig1a}b. The ordinate distance is from left to right along the white dashed line of Figure \ref{fig1a}b; the left-most arrow points to the brightenings before the minifilament formation and the two upward-pointing arrows point to the dark minifilament cross-section. (b) The negative flux as a function of time integrated over the white box of Figure \ref{fig1a}h. (c) HMI time-distance map of the pixels from left to right along the red dashed line of Figure \ref{fig1a}h. The blue and  green lines mark the minifilament formation time and jet eruption time, respectively.} \label{fig1b}
\end{figure} 

\section{INSTRUMENTATION AND DATA}\label{data} 

For our analysis we use EUV images obtained by AIA \citep{lem12} onboard \sdo\ in the 304, 171 and 193 \AA\ channels. AIA provides high spatial resolution of 0\arcsec.6 pixel$^{-1}$ ($\sim$ 430 km) with high temporal cadence of 12 s in seven EUV wavelength bands \citep{lem12}. Comparatively cool filament plasma appears dark in these EUV images against the coronal background of emission from the transition region and low corona because of the absorption of H and He photoionising radiation \citep{kuc98,anz05}. We primarily use 171 \AA\ images because we find pre-jet quiet-Sun minifilaments to be best seen in this channel.

We also use line-of-sight magnetograms from the HMI \citep{schou12} to observe the  photospheric magnetic flux in the minifilament and jet region. These magnetograms have high spatial resolution of 0\arcsec.5 pixel$^{-1}$ and temporal cadence of 45 s \citep{scherrer12}. 
With these magnetograms we follow the pre-jet evolution of the photospheric magnetic flux in the jet-base region, while examining nearly-concurrent EUV images of coronal emission. 

All of the jets were randomly found by using AIA images from JHelioviewer software\footnote{http://www.jhelioviewer.org}. The \sdo/AIA and \sdo/HMI data were downloaded from the JSOC cutout service\footnote{http://jsoc.stanford.edu/ajax/exportdata.html}, and co-aligned by using SolarSoft routines. Image alignment was checked by plotting HMI contours of active region magnetic patches onto the AIA images. For each jet, we examined data beginning at least 24 hours before jet eruption and we derotated all the images (AIA and HMI) to the time in the middle of the data set. For each event, we made AIA 5-min cadence moves, and HMI movie at about matching times, to study the minifilament formation process and the related photospheric magnetic field evolution. North is in the top and west is in the right of all the images and movies, and the x,y coordinates have their origin at disk center; x is positive (negative) west (east) of disk center, and y is positive (negative) north (south) of disk center.

We investigate the minifilament-formation preludes to the ten jet-eruption events that we studied in \cite{panesar16b}. That earlier work concentrated on the evolution of atmospheric magnetic structure and photospheric magnetic flux near the time of onset of the jets.  Here we examine longer time periods than we did in the \cite{panesar16b} work, so that we can understand the long-term build-up of the magnetic field of the minifilaments that eventually erupt to form the jets. It turns out that in several cases additional jets occur
on the same neutral lines as the jets listed in \cite{panesar16b}; these additional jets are essentially
homologous with those \cite{panesar16b} jets, and we include several of these additional jets in this study.  In order to maintain consistency with the \cite{panesar16b} work, the J1--J10 event numbers listed in Table 1 correspond to the exact same jets (same occurrence times and locations) as those in Table 1 of \cite{panesar16b}.  Our Table 1 here includes some of the homologous jets (date and time shown in italic font) listed with the jet of  \cite{panesar16b} that occurred on the same neutral line.  Thus for example, jet J9 of our Table 1 corresponds to jet J9 of Table 1 in \cite{panesar16b}, while the three additional jets listed under jet J9 here are homologous to J9 but were not included in the \cite{panesar16b} study.%, but occurred on the same neutral line as jet J9 in that earlier study.

We can identify three categories of jet eruptions in our table: (1) Single jet eruptions, whereby only a single significant jet occur over the entire lifetime of the neutral line on which the minifilament that erupts to make that jets resides. Jets J1, J2, J3, J5, and J8 in Table 1 are of this category. (2) Sequences of jets that are clearly homologous, in the sense that the jets originate from the same neutral line, and the minifilament that erupts to make these jets becomes totally undetectable for a period of time between successive jets.  This temporary disappearance of the minifilament means one of two things: either the minifilament erupted completely to form a jet, and then a new minifilament reforms/reappears and later erupts again to form a second jet; or in the earlier eruption only a portion of the minifilament erupts, with the remaining portion of the minifilament becoming temporarily invisible due to obscuration of surrounding hot and/or cool material, and with another portion of that remaining minifilament erupting sometime later to form the subsequent jet. Sequential jets in this category  tend to be separated in time by several hours, which is consistent with the minifilaments totally erupting and then reforming anew in-between jet events. Nonetheless, we do not belabor trying to discern among these two possibilities (total minifilament eruption in-between jets, or partial minifilament eruption with temporary obscuration of the remaining portion), but instead just point out that the minifilament reforms or reappears in-between successive eruptions. We have two sets of examples of such clearly homologous series of jets in our data set, with one set being the four jets
listed with J9, and the second set consisting of the pair J6 and J7; these latter two were not discussed in terms of homology in \cite{panesar16b},  which is why they have separate numbers.  But we now recognize that these two jets originated from the same neutral line, and are therefore homologous. (3) Cases where successive jets clearly originate from successive partial eruptions of a single minifilament. That is, first one part of the filament erupts, causing a jet, and then the remaining part of the filament
erupts to form the second jet. In these cases, the time between successive jets is only 30-40 min. In fact, it is not clear whether these ``pairs'' of jets should be referred to as a single jet, that is, as two episodes of the same jet eruption.  We also do not concern ourselves with this largely-semantic question. Jets J4 and J10 of Table 1 belong to this category.

\begin{figure*}
	\centering
	\includegraphics[width=\linewidth]{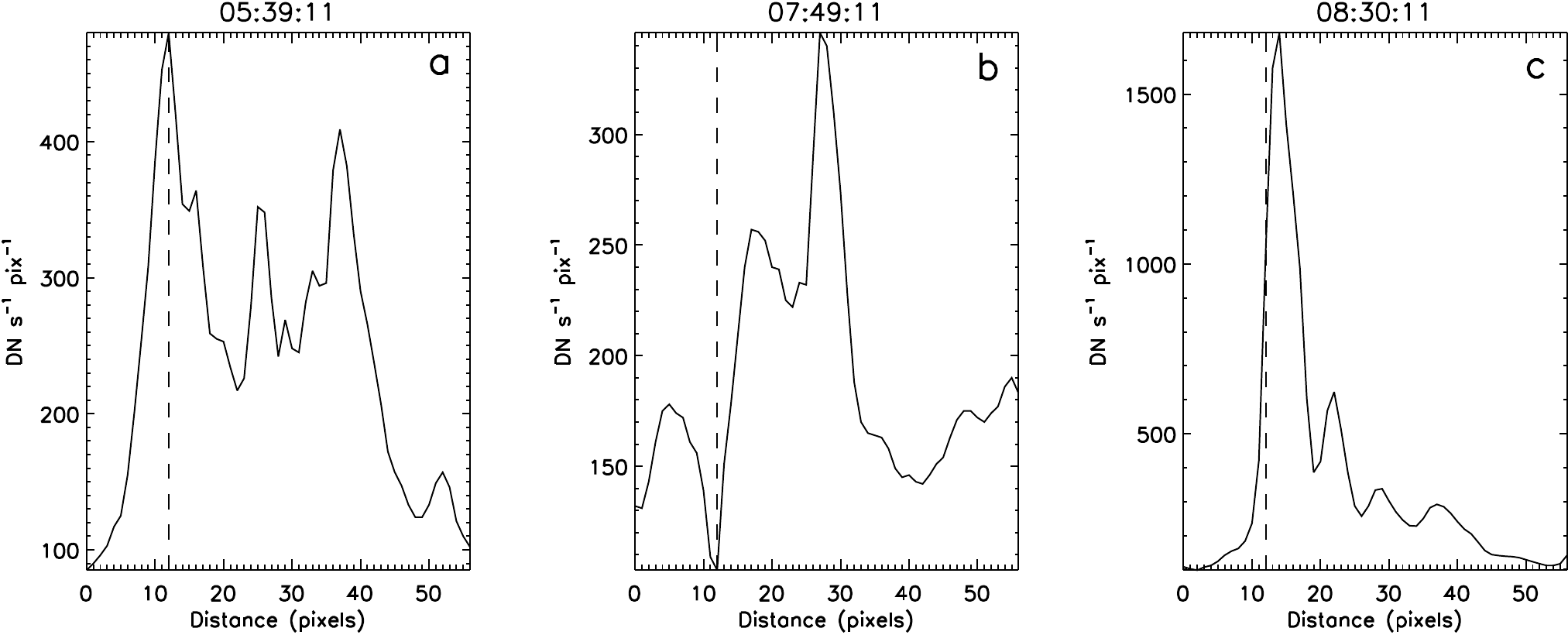}
	\caption{Plots of the 171 \AA\ intensity through the location of pre-jet minifilament (J2)  from left to right along the white dashed line of Figure \ref{fig1a}b. These plots are at the times of the 171 \AA\ panels of Figure \ref{fig1a}; (a) before  minifilament formation, (b) during the presence of the minifilament, and (c) at the time of minifilament-eruption onset.  In each panel, the vertical black dashed line shows the location of where the minifilament resides in panel (b).} \label{fig1c}
\end{figure*}

\begin{figure*}
	\centering
	\includegraphics[width=\linewidth]{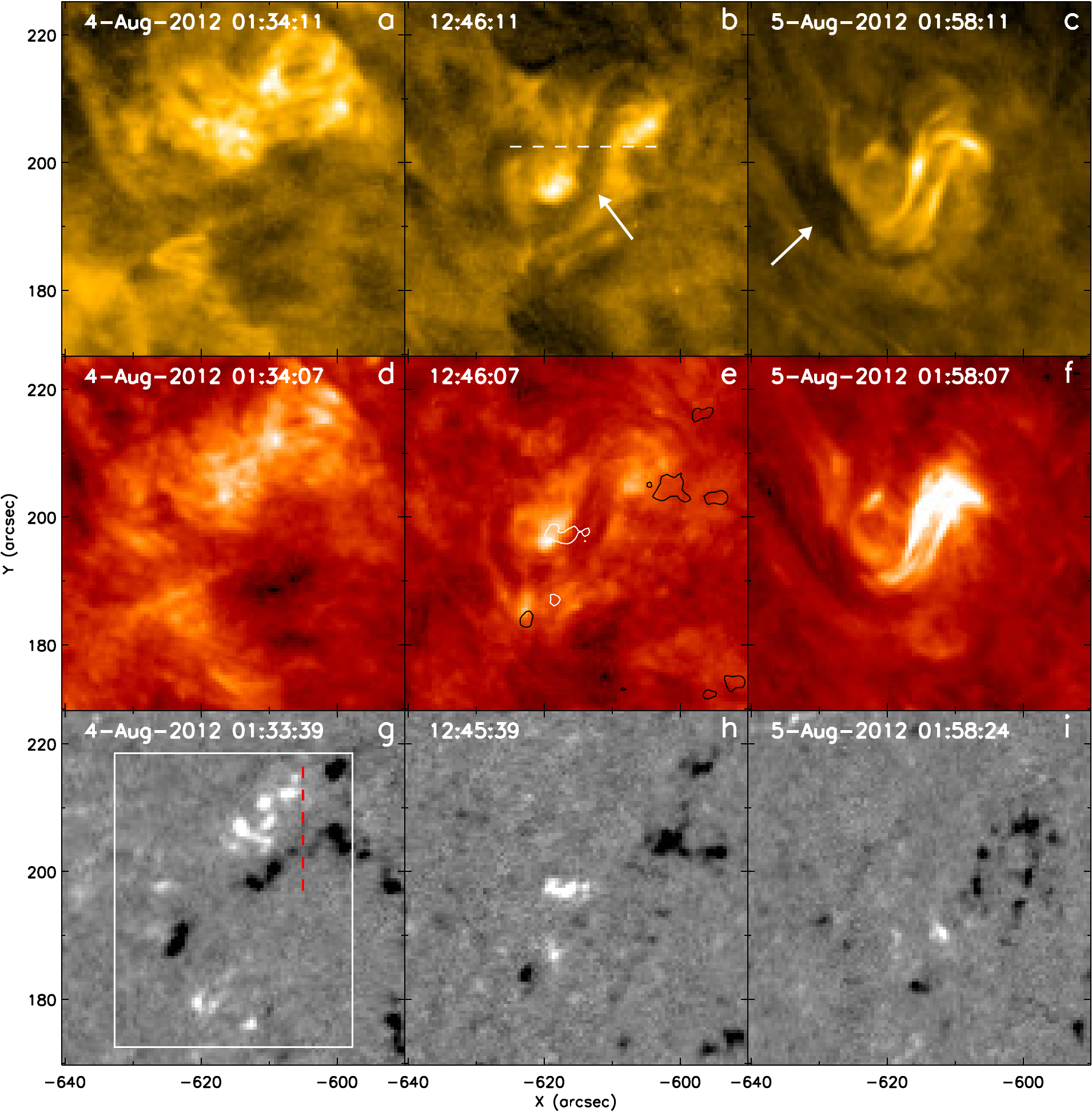} 
	\caption{Region of the quiet-region minifilament that led to jet J4 of Table \ref{tab:list}, showing (a-c) 171 \AA\ and (d-f)
		304 \AA\ intensity images from AIA, and (g-i) magnetograms of the same region from HMI\@.  Dates at the top of each column apply to each of the three panels in those respective columns.  Left, middle, and right columns respectively show the region before minifilament formation, a time when the minifilament is present, and a time when the minifilament is erupting to form the jet. In (b), the arrow points to the minifilament and the white dashed line shows the east-west cut for the time-distance map of Figure \ref{fig2b}a. The arrow in (c) points to the cool minifilament material during eruption onset. The  boxed region in (g) shows the area used for measuring the positive magnetic flux for the plot shown in  Figure \ref{fig2b}b; the red dashed line shows the north-south cut for the Figure \ref{fig2b}c time-distance map. HMI contours (level $\pm$ 50 G) of (h)  are overlaid onto panel (e), where white and black respectively represent positive and negative
		polarities. 
		Animations (MOVIE2) of this Figure are available.} \label{fig2a}
\end{figure*}

%\vspace{0.3cm}
%
\section{RESULTS}\label{result}
\subsection{\textit{Overview}}
% All the ten events are listed in Table \ref{tab:list}.
We carried out a detailed analysis of the formation and evolution of each of the minifilaments that trigger the coronal jet eruptions of Table \ref{tab:list}. For the detailed analysis, first we follow the minifilament back in time in the EUV images to find when it is born. Then we study the evolution of the minifilament up to its eruption. Similarly, we track the photospheric field evolution of the minifilament region.  In the following we present in detail the origin of three minifilaments (J2, J4, and J7 from Table \ref{tab:list}): We examine their formation and evolution primarily in EUV images in  Section \ref{evo1}, and we concentrate on the photospheric magnetic field evolution in Section \ref{flux}. 

\begin{figure}
	\centering
	\includegraphics[width=\linewidth]{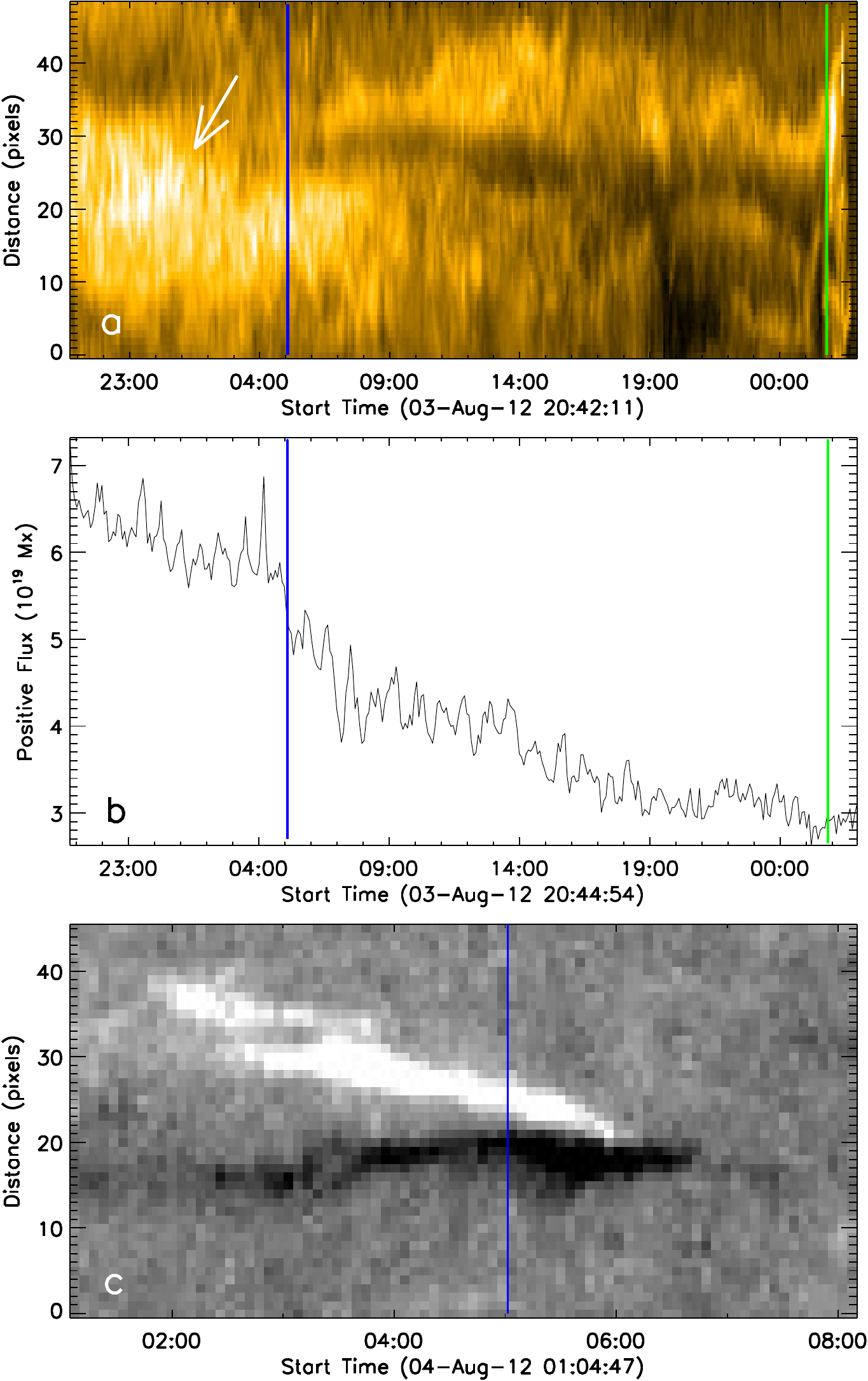}
	\caption{Minifilament formation and flux cancelation for jet J4: (a) 171 \AA\ time-distance map of the pixels from left to right along the white dashed line of Figure \ref{fig2a}b; the white arrow points to the brightenings before the minifilament formation. (b) The positive flux as a function of time integrated over the white box of Figure \ref{fig2a}g. (c) HMI time-distance map of the pixels from bottom to top along the red dashed line of Figure \ref{fig2a}g. It shows the flux evolution during the minifilament formation time. The blue line and the green line indicate the minifilament formation time and jet eruption time, respectively.} \label{fig2b}
\end{figure} 

\subsection{\textit{Formation and Evolution of Minifilaments}}\label{evo1}

Figure \ref{fig1a} shows the jet J2 (Table \ref{tab:list}) source location in AIA 171 \AA\ and 304 \AA\ EUV images before minifilament formation (Figures \ref{fig1a}a and \ref{fig1a}d), at a time when the minifilament is present but has not yet begun to erupt (Figures \ref{fig1a}b and \ref{fig1a}e), and at the time of the minifilament's eruption onset (Figures \ref{fig1a}c and \ref{fig1a}f). The minifilament forms in the bright pre-jet region in Figures \ref{fig1a}a,b,d,e, and this bright region becomes the base of the jet in Figures \ref{fig1a}c,f. We also include an accompanying movie  (MOVIE1), which shows the inspected region for $\sim$ 12 hours. 
When the minifilament starts to form, we observe enhanced brightenings at the neutral line (e.g.\@ at 05:39 UT in Figures \ref{fig1a}a and \ref{fig1b}a), and shortly thereafter (at about 05:58 UT in MOVIE1) the minifilament appears above the neutral line (Figure \ref{fig1a}e). The minifilament steadily gets darker and more prominent (shown with an arrow in Figure \ref{fig1a}b), eventually achieving a length of 25,000 km (from Table 1 of \citealt{panesar16b}) and width of 1500 km. It persists for 2.5 hours before erupting. The JBP appears at the minifilament neural line during the minifilament eruption, starting at 08:29 (arrow in Figure \ref{fig1a}c). 

We also observe some  dark features in the pre-jet region (black arrows in Figure \ref{fig1a}a) that are nearly normal to the neutral line of two diverging flux clumps (Figure \ref{fig1a}g); these are \textit{field transition arches} of emerging field loops \citep{bruzek77}.

To examine the minifilament's formation and evolution in detail, we created a time-distance map of the region for 4.5 hours, covering a period before and during the formation. 
Figure \ref{fig1b}a shows the evolution of the minifilament in the cut of the white dashed line in Figure \ref{fig1a}b. The ordinate distance is from left to right along the white dashed line of Figure \ref{fig1a}b. 
The blue line in Figure \ref{fig1b}a marks the time when the cool-transition-region minifilament plasma starts appearing in the AIA 171 \AA\ images. As mentioned above, first we observe brightenings at the location where the filament subsequently forms (white arrow in Figure \ref{fig1b}a). We can see that the minifilament becomes clearly visible (see arrows in Figure \ref{fig1b}a) and remained stable for 2.5 hours after its formation. The minifilament shows motions from at least 08:22 UT, and it is rapidly rising by 08:28 UT as it begins erupting and making the jet (white arrow in Figure \ref{fig1a}c, and immediately to the left of the green line in Figure \ref{fig1b}a). It first shows early motions consistent with a slower-rise phase from 08:22 UT to about 08:28 UT, followed by motions consistent with a faster-rise phase from about 08:28 UT, and then it becomes part of the jet spire by 08:32 UT. A similar progression occurs in all the ten events (details of the jet eruptions  are given in \citealt{panesar16b}) as well as in at least some active region jets \citep{sterling16}. A similar slow-rise/fast-rise is also typically seen in larger-scale filament eruptions \citep[e.g.][]{tand80,sterling05,panesar15}.

In Figure \ref{fig1c}, we display 171 \AA\ intensity plotted from left to right along the white dashed line in Figure \ref{fig1a}b, Figures \ref{fig1c}a,b,c are respectively at the same times as Figures \ref{fig1a}a,b,c. The black dashed lines in Figure \ref{fig1c} show the intensity at the location of the minifilament along the white dashed line of Figure \ref{fig1a}b. Figure \ref{fig1c}a shows the above-mentioned increased  intensity at the location of the minifilament from a few minutes before its formation.  There is a strong dip in the intensity in Figure \ref{fig1c}b at the location of the dashed line, which is due to the presence of the minifilament cool-transition-region plasma at that location after the minifilament forms. During the eruption onset (Figure \ref{fig1c}c), the intensity at that location has again increased because of the appearance of the JBP at the location where the pre-eruption minifilament resided.

\begin{figure*}
	\centering
	\includegraphics[width=\linewidth]{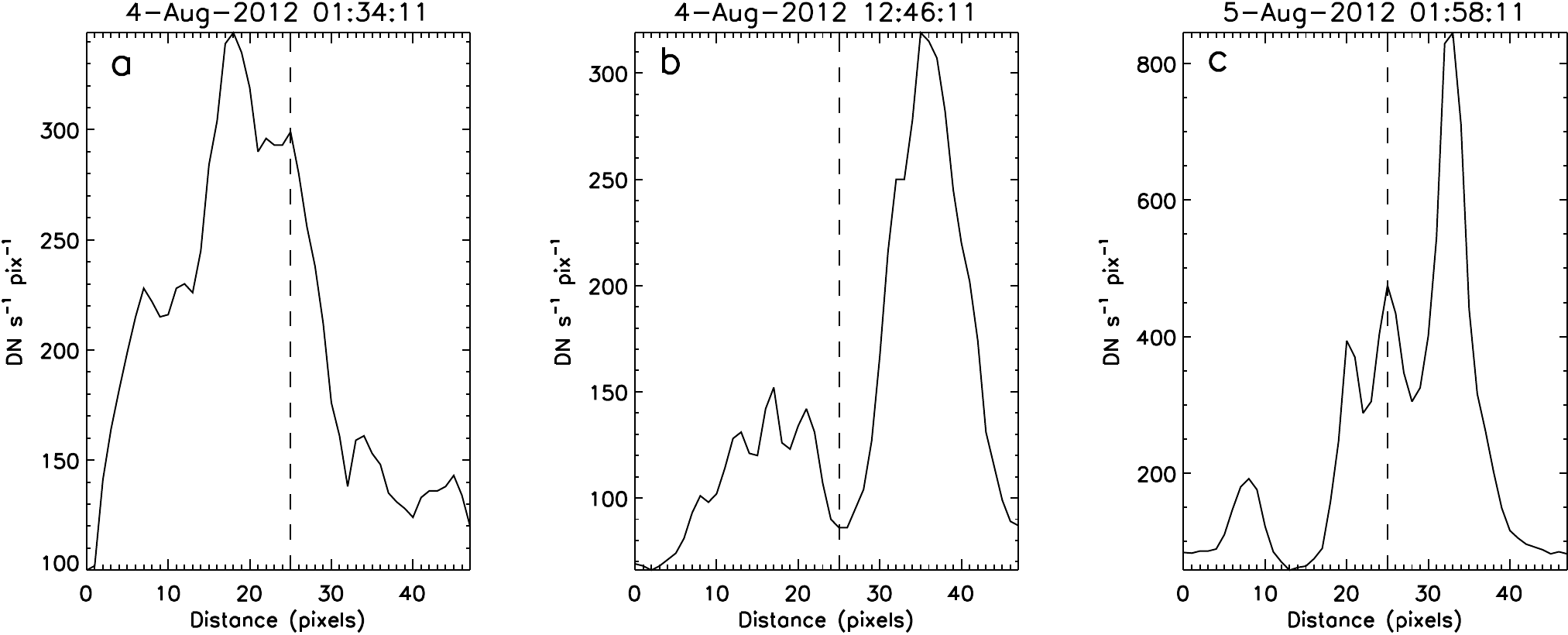}
	\caption{Plots of the 171\AA\ intensity through the location of the  pre-jet minifilament (J4)  from left to right along the white dashed line of Figure \ref{fig2a}b. These plots are at the times of the 171 \AA\ panels of Figure \ref{fig2a}; (a) before  minifilament formation, (b) during the presence of the minifilament, and (c) at the time of minifilament-eruption onset. In each panel, the black dashed line shows the position of where the minifilament resides in panel (b).} \label{fig2c}
\end{figure*}

In Figure \ref{fig2a}, we show another example of a jet (J4) from our list. We analyzed the region for $\sim$ 39 hours; see MOVIE2. Figure \ref{fig2a}a shows the region before the formation of the minifilament. The minifilament subsequently appears above a neutral line at $\sim$ 05:04 UT on 2012 Aug 04 (MOVIE2). As it starts to form, initially the filament body is weak and only visible in fragments instead of at a complete continuous structure. Later at about 12:56 UT, a definite minifilament  develops (Figures \ref{fig2a}b,e) with a length and width of 31,000 km and 2500 km, respectively (Table 1 of \citealt{panesar16b}). The minifilament then was stable for about 21 hours. The minifilament starts to lift-off at 01:58 UT and then the JBP appears at the neutral line (Figure \ref{fig2a}c) where the minifilament resided prior to the eruption (MOVIE2). After the appearance of the JBP, the jet spire starts to extend upward.

In Figure \ref{fig2b}, we present plots for this event similar to Figure \ref{fig1b} for the previous event.  Figure \ref{fig2b}a shows a 30-hr time-distance map of the pixels along the white dashed line of Figure \ref{fig2a}b. The white arrow points to brightenings prior to minifilament formation occurring at the location of the neutral line (Figure \ref{fig2a}g) on which the minifilament later forms. The minifilament forms  at $\sim$ 05:04 UT on 2012 Aug 04 (shown by the blue line in Figure \ref{fig2b}a) as the opposite-polarity flux patches were canceling (this is consistent with the observations and models of typical filament formation, e.g.\@ \citealt{martin85,Balle89,martens01}). Even before that time we can see at that location some signatures of transient filamentary-type  cool material, which suggests that the minifilament is in its formation stage. During the formation process the minifilament first appears as a weak and faint structure, which disappears for sometime and then reappears. (This, in fact, is common to all ten events.) There are two nearly-overlapping eruptions in this region, with the first one starting  at 01:58 UT and the second at 02:20 UT; we will discuss further this in Section \ref{homo}. The green line in Figure \ref{fig2b}a shows the time (01:58 UT on 2012 Aug 05) when the minifilament starts to rise (we do not add a separate line for the start of the final eruption, which started within a half  hour of the green line).

\begin{figure*}
	\centering
	\includegraphics[width=0.85\linewidth]{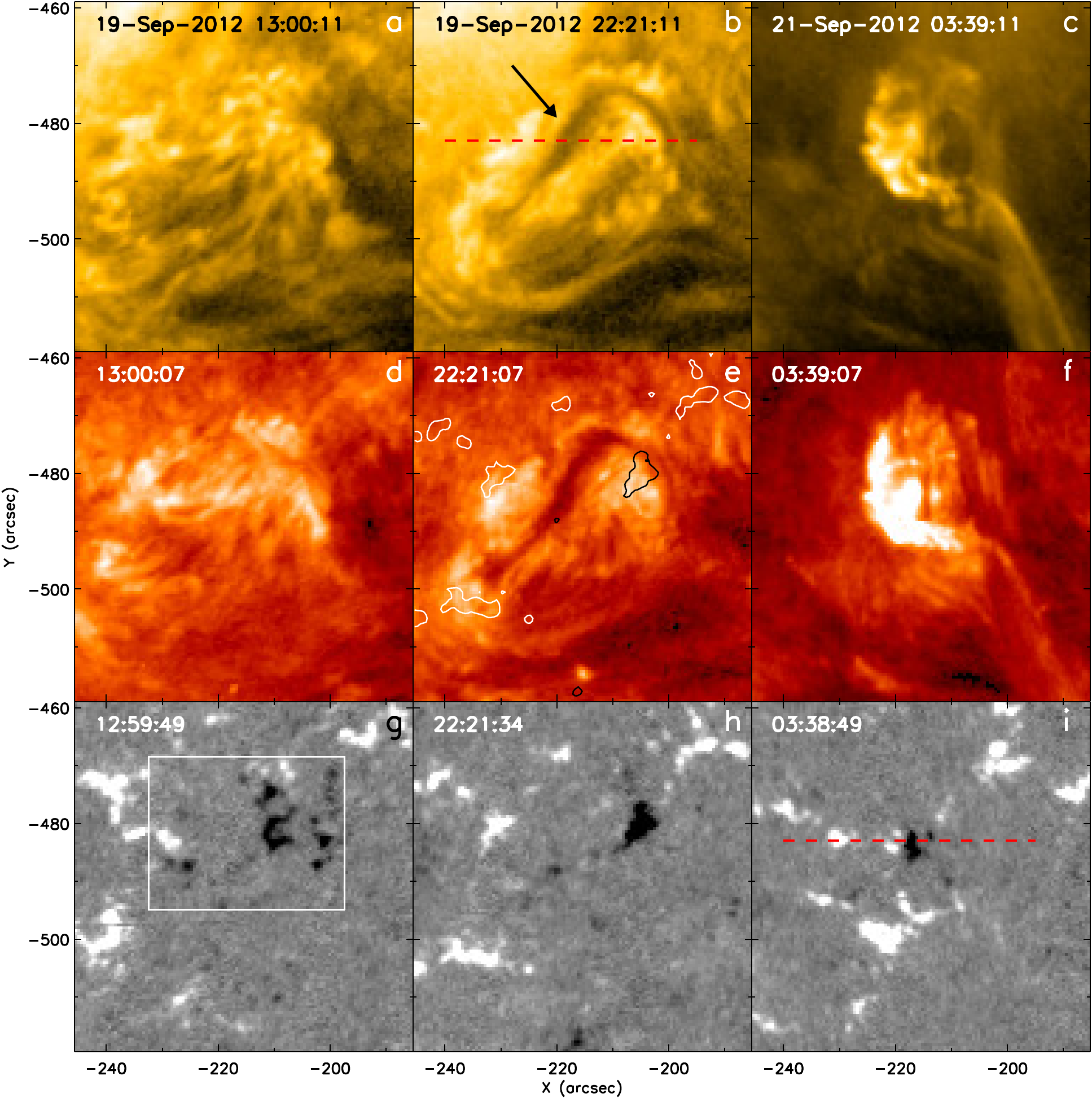}
	\caption{Region of the quiet-region minifilament that led to jet J7 of Table \ref{tab:list}, showing (a-c) 171 \AA\ and (d-f)
		304 \AA\ intensity images from AIA, and (g-i) magnetograms of the same region from HMI\@.  Dates at the top of each column apply to each of the three panels in those respective columns.  Left, middle, and right columns respectively show the region before minifilament formation, a time when the minifilament is present, and a time when the minifilament is erupting to form the jet. In (b), the arrow points to the minifilament and the red dashed line shows the east-west cut for the time-distance map of Figure \ref{fig3b}a. The  boxed region in (g) shows the area used for measuring the negative magnetic flux for the plot shown in  Figure \ref{fig3b}b; the red dashed line in (i) shows the east-west cut for the Figure \ref{fig3b}c time-distance map. HMI contours (level $\pm$ 50 G) of (h)  are overlaid onto panel (e), where white and black respectively represent positive and negative polarities. 
		Animations (MOVIE3) of this Figure are available.} \label{fig3a} 
	
\end{figure*} 

In Figure \ref{fig2c}, we present plots of the 171 \AA\ intensity from left to right along the  white dashed line of Figure \ref{fig2a}b, at three times in the evolution of the pre-jet region. Figures \ref{fig2c}a and \ref{fig2c}c show the increased intensity on and near the neutral line just prior to formation of the minifilament and during the eruption onset of the minifilament, respectively, just as in Figure \ref{fig1c} for the previously-discussed event. In Figure \ref{fig2c}b, the intensity dips at the location of the cool-transition-region-plasma minifilament (marked by the dashed black line in Figure \ref{fig2c}b). %Again intensity increases when the minifilament erupts and JBP appears at the location of minifilament (dashed line in Figure \ref{fig2a}).

Similarly, in Figure \ref{fig3a}, we show the third example from our list, J7 of Table \ref{tab:list}. In this case there were two jet eruptions (J6 and J7) from the same neutral line, about 5 hours apart. In total, we studied this region for $\sim$ 39 hours (MOVIE3), again starting before the formation of the minifilament, and lasting through the two minifilament eruptions and corresponding jets. The minifilament forms at 17:15 UT on 2012 Sep 19, about 30 hours before its first jet eruption. It erupts outward to form jet J6 at 22:52 UT on 2012 Sep 20 (see MOVIE3). After the jet eruption (J6), the minifilament slowly starts to reappear/reform at the same neutral line, and again erupts at 03:33 UT on 2012 Sep 21 to drive jet J7 (Figure \ref{fig3a}c). 
%There is some bright haze (in MOVIE3) in the north of the minifilament, which is due to the plume (not shown in the images). 

For this sequence of events, Figure \ref{fig3b}a shows the time-distance map from left to right along the red dashed line of Figure \ref{fig3a}b. The blue line in Figure \ref{fig3b}a marks the time when the minifilament structure becomes clearly visible; prior to that time, the minifilament structure was weak, and  not continuously visible. With time it became darker and more stable (lasting for 39 hours). As mentioned above first it erupts to make jet J6, and after reforming it again erupts to drive jet J7, which is the final jet (see the brightenings just after the green lines in Figure \ref{fig3b}a) in that homologous series. 

In Figure \ref{fig3c}, we show 171 \AA\  intensity plots, from left to right along the red dashed line of Figure \ref{fig3a}b. The intensity peaks across the neutral line prior to the minifilament's formation and after the minifilament's eruption (e.g. in Figures \ref{fig3c}a and \ref{fig3c}c), and the intensity there is very low during the presence of minifilament (Figure \ref{fig3c}b).%The brightenings at the edges of the minifilament in Figure \ref{fig3a}b contributes to the peak in intensity in Figure \ref{fig3c}b. 

\subsection{\textit{Magnetic Flux Cancelation}}\label{flux}

%To investigate the relationship between the photospheric magnetic field and minifilament/jet regions, we inspect the HMI line-of-sight magnetograms for all our events listed in Table \ref{tab:list}. 

Figures \ref{fig1a}g-i display the photospheric magnetic field environment for jet J2 before the minifilament formed, during a period when the minifilament existed in a stable state, and at the time of the eruption onset, respectively.  There is an emergence of a new bipole at $\sim$ 22:30 UT on 2012 Jun 30 (MOVIE1), and the positive-flux foot of the newly-emerged bipole soon begins canceling with  pre-existing negative flux, which results in the formation of the minifilament at 05:58 UT on 2012 Jul 01 at the cancelation site.  The minifilament forms along the neutral line between the minority-polarity positive-flux patch of the emerged bipole, and majority-polarity pre-existing negative-flux patch (Figure \ref{fig1a}h). MOVIE1 accompanying Figure \ref{fig1a} shows the complete evolution of the flux, including flux emergence in the beginning and then cancelation leading to the minifilament formation and the jet J2 eruption. 

To examine and follow the magnetic flux evolution quantitatively with respect to the minifilament evolution, we measured the negative flux (minority-polarity flux) inside the white box of Figure \ref{fig1a}h. Here, and with all of our flux measurements, we were careful to ensure that the boxed flux was well isolated, in order to minimize flows of the selected polarity (positive polarity in this case) across the boundaries of the box. Figure \ref{fig1b}b shows the integrated flux as a function of time. A  continuous decrease in the negative flux can be seen, which is clear evidence of flux cancelation at the neutral line. During the flux-cancelation, the minifilament forms at the time marked by the blue line in Figure \ref{fig1b}. The brightenings that occurs before the minifilament formation can be seen at the neutral line between 05:30 and 06:00 in MOVIE1 and in Figure \ref{fig1b}a. 

To see more clearly the flux convergence and cancelation, Figure \ref{fig1b}c shows a time distance map along the horizontal red dashed line of Figure \ref{fig1a}h. (Note: In Figure \ref{fig1b}c we plotted flux for 13.5 hours whereas in Figures \ref{fig1b}a,b the plots cover only 4.5 hours.) A positive patch of the newly-emerged bipole migrates to the south-east and starts canceling with nearby negative flux (Figure \ref{fig1a}g). First we observe brightenings at the cancelation site, and then the minifilament forms; thus this initial cancelation seems to explain the pre-minifilament-formation brightening that we see in the three events presented in this paper. As more and more flux cancels, the dark filament body becomes more prominent and thick. Positive and negative flux patches (Figure \ref{fig1b}c) are seen to converge and cancel at the neutral line even after the minifilament formation. (Flux cancelation is not obvious between the blue and green lines in Figure \ref{fig1b}c because of the selection of our cut location, shown by the line in Figure \ref{fig1a}h; for the time between the blue and green lines cancelation does not occur exactly on that Figure \ref{fig1a}h line. Cancelation is occurring however at nearby locations that are included in the box of Figure \ref{fig1a}g. That is why the flux decrease is obvious between the blue and green lines in Figure \ref{fig1b}b, but not in Figure \ref{fig1b}c.) It appears that the continuing flux cancelation at the neutral line finally destabilizes the field holding the minifilament, and that field then erupts to make the jet (at the time marked by the vertical green line in Figure \ref{fig1b}). This is in agreement with the flux rope model of solar filaments, discussed by \cite{Balle89}, where  progressive reconnection at the neutral line beneath the filament builds and then destabilizes the field holding the filament material, resulting in the filament eruption.

\begin{figure}
	\centering
	\includegraphics[width=\linewidth]{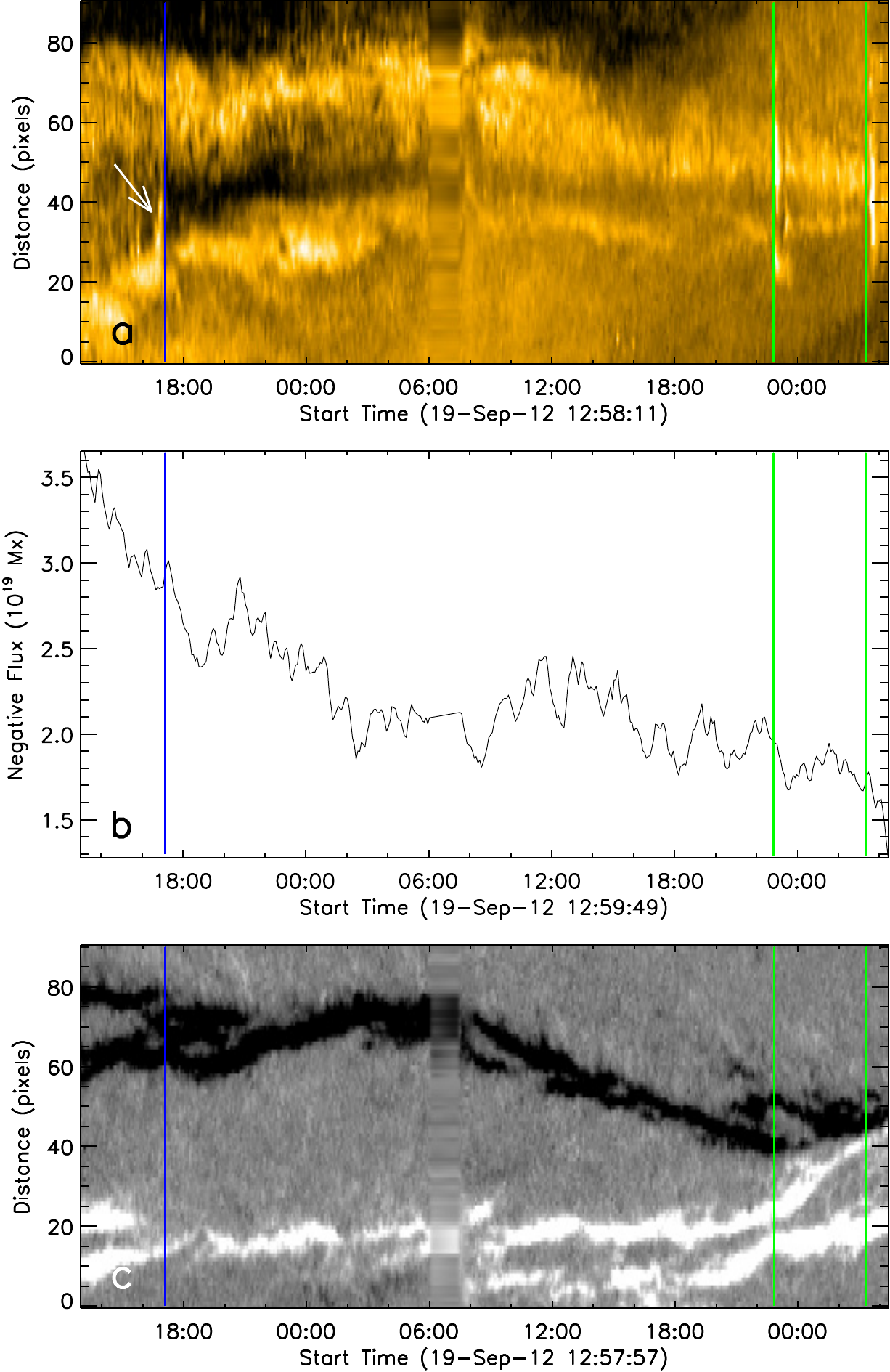}
	\caption{Minifilament formation and flux cancelation for jets J6 and J7: (a) 171 \AA\  time-distance map of the pixels from left to right along the red dashed line of Figure \ref{fig3a}b; the white arrow points to the brightenings on the neutral line as the minifilament forms. (b) The negative flux as a function of time integrated over the white box of Figure \ref{fig3a}g. (c) The HMI time-distance map of the pixels from left to right along the red dashed line of Figure \ref{fig3a}i. The blue line indicates the minifilament formation time and the two green lines indicate the jet eruption times for J6 and J7.} \label{fig3b}
\end{figure}

\begin{figure*}
	\centering
	\includegraphics[width=\linewidth]{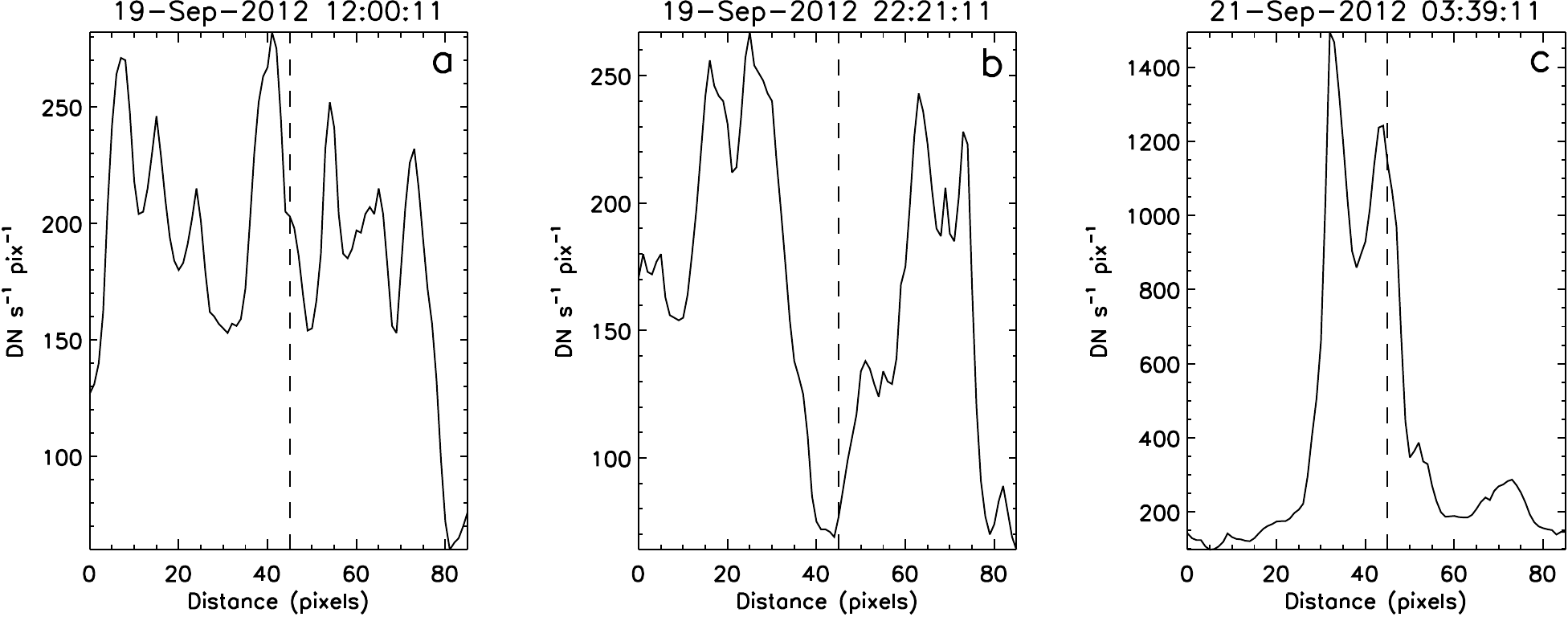}
	\caption{Plots of 171 \AA\ intensity through the location of the pre-jet minifilament (J7) from left to right along the red dashed line of Figure \ref{fig3a}b. These plots are at the times of the 171 \AA\ panels in Figure \ref{fig3a}; (a) before  minifilament formation, (b) during the presence of the minifilament, and (c) at the time of minifilament eruption onset. In each panel, the black dashed line shows the position of where the minifilament resides in panel (b).} \label{fig3c}
\end{figure*}

We notice similar type of magnetic behavior in jet J4 (Figures \ref{fig2a}g-i). First, we observe brightenings at the cancelation site (white arrow in Figure \ref{fig2a}), and then the minifilament slowly starts to appear in sheared field above the neutral line.%, e.g. at 12:04 UT on 2012 Aug 04 in MOVIE2.

For the same jet J4, in Figure \ref{fig2b}b we show the integrated positive flux as a function of time, for $\sim$ 30 hours, over the white box of Figure \ref{fig2a}g. We observe a continuous decrease in positive flux throughout the time period, but the minifilament becomes apparent in the EUV images at the time of a precipitous drop in flux (blue line in Figure \ref{fig2b}b). The jet J4 occurs at the time indicated by the green line in Figure \ref{fig2b}, as flux cancelation continued at the neutral line (Figure \ref{fig2b}b). Figure \ref{fig2b}c shows a time-distance map from the red dashed line of Figure \ref{fig2a}g. (Note: Figure \ref{fig2b}c shows the time-distance map for $\sim$ 7 hours whereas Figures \ref{fig2b}a,b are for $\sim$ 30 hours.) It displays strong cancelation at the time of minifilament formation (marked by the blue line). %This image shows the cancelation only at the north of the neutral line, rest of the dynamic changes in the flux can be seen in the MOVIE2.

Figures \ref{fig3a}g-i show the magnetic setup for jets J6 and J7. The minifilament lies along the neutral line that runs between the majority-polarity flux (positive) and minority-polarity flux (negative) (Figures \ref{fig3a}e,h). 

In Figure \ref{fig3b}b, we plot the negative flux values integrated over the white box region of Figure \ref{fig3a}g. We observe  long-term flux cancelation for about two days in the pre-jet region. We can see similar behavior of the flux in the time-distance map (Figure \ref{fig3b}c, also in MOVIE3), which is plotted from the red dashed line of Figure \ref{fig3a}i. The opposite-polarity flux patches approach each other, and flux cancelation between them leads to the filament formation (at the time of the blue line in Figure \ref{fig3b}). In addition to these relatively large flux patches, there are also some small flux grains close to the neutral line that canceled throughout the time period, e.g. between 17:00 and 19:00 UT on 2012 Sep 19 in MOVIE3. Cancelation of the large flux patches as well as cancelation of flux grains continues through the times of the two coronal jets (marked by the green lines in Figure \ref{fig3b}). Thus, our flux plots and time-distance maps confirm that the minifilament formation, its steady life, and its eruption all occur in conjunction with flux cancelation.

Similarly, we tracked and compared the magnetic field evolution with the minifilament formation and evolution for the remaining events of Table \ref{tab:list}. We found similarly clear evidence for flux cancelation in all  of the events. We also estimated the reduction  of the minority-polarity flux from before to after the minifilament formation (see Table \ref{tab:list}). In all cases, 10-40\% of the flux (although the percentage for event J10 is with the 1$\sigma$ uncertainty) canceled over a 3-4 hour time period around the time of minifilament formation (measured from $\sim$ 3 hours prior to minifilament formation to $\sim$ 1 hour after formation).

\subsubsection{\textit{Homologous and/or Partial Jet Eruptions}}\label{homo}

In four of our events (J4, J6, J9 and J10; Table \ref{tab:list}), we observe more than a single jet from the same neutral line; that is, they are homologous, in the sense that they originate from the same location (same neutral line) and have nearly identical structure and evolution \citep{dodson77}. What accounts for the production of multiple jets from the same neutral line? We find that recurrent minifilament eruptions drive these homologous jets. A minifilament erupts and drives a jet, reforms/reappears at the same location, and then again erupts, driving the next jet. This process occurs as flux cancelation is ongoing, and continues until all the minority-polarity flux vanishes. For example in jet J6, a minifilament erupts at 22:52 UT on 2012 Sep 20 and reforms/reappears at 00:51 UT on 2012 Sep 21 (see MOVIE3). From concurrent magnetograms we can see that a significant amount of flux still remains after the first jet, and that flux  continues to cancel and leads to subsequent minifilaments and jet eruptions. For jet J9, we find four homologous jet eruptions within a two-day period from the same neutral line; comparison with concurrent magnetograms indicates that this too is due to continuous flux cancelation at the neutral line; the jetting stops when the minority flux has completely disappeared. 

Jets J4 and J10 also display multiple eruptions, but these cases are slightly different from the above-mentioned cases. For these cases, we observe a cascade of eruptions from the same neutral line within a 30-40 minute time period. The only difference is that one part of the minifilament erupts first to make a jet, which is then followed by a final jet eruption. The minifilament structures are seen to erupt completely during the final eruptions. 
%In two jets J4 and J10, we observe the partial minifilament eruptions; one part of the minifilament erupts first to produce a jet, and within 10-15 minutes the  minifilament completely erupts to form a final jet in the sequence. 

In all ten events, we find that flux cancelation is the basic process for the formation of the minifilaments and for triggering their jet-driving eruptions.

\section{SUMMARY AND DISCUSSION}\label{discussion}

In \cite{panesar16b} we studied ten on-disk quiet-region coronal jets, and found that they were driven by minifilament eruptions following flux cancelation at the neutral line on which the  minifilaments resided. Here we investigated the magnetic origin and/or formation of pre-jet minifilaments for the same ten events of \cite{panesar16b}. We found that each of the ten pre-jet minifilaments  formed due to flux cancelation between majority-polarity and minority-polarity flux clumps.  In all of the events, the minifilament forms in highly sheared field above the  neutral line along which the cancelation was occurring. The flux  continued to cancel even after the formation of the minifilaments and this eventually led to the minifilament eruption and the resulting jet. In each case the flux cancelation continued until the minority-polarity flux patch completely disappeared. 

For each event, we followed the HMI data backwards in time to learn the origin of the flux clumps.  We found the following three scenarios: (a) In four events (J3, J5, J8 and J10), tiny grains of flux coalesce to make a minority-polarity flux clump; (b) in two events (J2 and J9) the minority-polarity foot of a newly-emerged bipole became the minority-polarity flux clump; (c) in four events (J1, J4, J6 and J7), the minority-polarity clump preexisted as it rotated onto the Earthward side of the Sun 2-3 days before our observations began. In each case, the minority-polarity flux clump migrated toward the majority-polarity flux clump for cancelation that resulted in the formation of the minifilament. 

Cancelations that include relatively large minority-polarity flux clumps can make homologous jet eruptions: a minifilament forms, remains stable for a period of time, and then erupts to form a jet; a minifilament then reforms/reappears on the cancelation neutral line and eventually again erupts and forms a new jet. This process continues as long as the cancelation of the minority-polarity flux patch continues (e.g. in J6 and J9). Eventually, the minority-polarity flux patch disappears, and thus the polarity inversion line disappears and jetting stops. Apparently, the jetting stops when the condition for the existence of a minifilament is no longer present, and that condition is apparently on-going flux cancelation at a neutral line. Similarly, \cite{sterling17} find jetting in active region jets to  continue only as long as flux cancelation is occurring. 

Our broader studies show that there are some differences, in terms of minifilaments, 
between jets in quiet Sun and coronal holes and jets in active regions.  While we usually see minifilaments erupting to form jets in CHs  \citep{sterling15} and in quiet Sun \citep{panesar16b}, minifilaments that erupt to form jets in active regions are sometimes hard to detect \citep{sterling16,sterling17}.  We nonetheless suspect that magnetic eruptions similar to minifilament eruptions also drive most or all such active region jets, but the erupting minifilaments in active regions often appear narrower (thread-like structures) than in jets in quiet regions or coronal holes; erupting minifilaments in active regions tend to be obscured by surrounding bright and dark features; and cool minifilament material may be absent in the pre-erupting field of some active region jets. \cite{sterling16,sterling17} examine the similarities/differences of active region jets and non-active region jets in more detail.
% The QS jets studied in this paper and in \cite{panesar16b}, the CH jets studied in \citep{sterling15}, and at least
%	some the AR jets in \citep{sterling16,sterling17} are consistent with beginning in basically the same way: by eruption of a twisted flux rope (the minifilament field) that is made and triggered to erupt by flux cancelation at the PIL of a small magnetic arcade.
% 

It has been proposed that typical filaments form due to a flux-linkage process  \citep[e.g.][]{martens01,panesar14c}. According to the flux-linkage model, two initially unconnected bipoles undergo magnetic reconnection with each other due to flux convergence and cancelation of photospheric magnetic flux, and this results in the formation and evolution of a filament. The convergence of opposite-polarity flux patches towards a neutral line has commonly been observed before the formation of filaments, and is a basic aspect of various theoretical models for filament formation in addition to flux linkage (e.g. \citealt{litvinenko99,gai97,devore00,Balle89,balle90,balle00}). Therefore, magnetic flux cancelation seems to play an important role in the formation of typical filaments \citep{martin85,martin86,mar98,tand95}. Small-scale filaments have also been observed to form at sites of cancelation \citep[e.g.][]{hermans86,wang00,lees03}. Our results here are consistent with these earlier findings, as we find that  minifilaments form when two initially unconnected (or relatively poorly connected) magnetic flux patches undergo reconnection and magnetic flux cancelation. Continued flux cancelation at the neutral line leads to  the minifilament eruption that drives a jet.
%Our minifilaments also form and erupt due to the magnetic flux cancelation at the neutal line.
\begin{figure*}
	\centering
	\includegraphics[width=\linewidth]{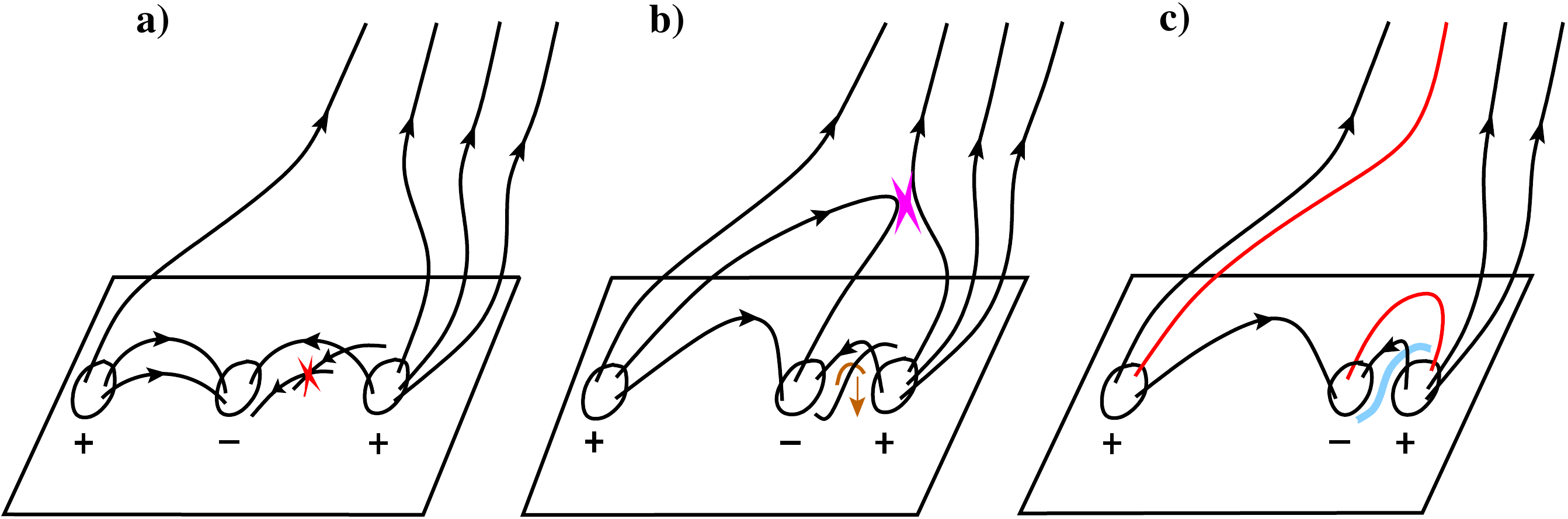}
	\caption{Schematic interpretation of the observations: The curved lines represent magnetic field lines, some of which form closed loops, and others of which are either open or far-reaching-closed fields (AIA images for our events show that ambient coronal field in the region of jets is often a far-reaching field instead of truly open). A minority-polarity (negative-polarity black oval) flux clump sits in-between two majority-polarity (positive-polarity black ovals) flux clumps. The two right-most flux clumps converge over time (progressively from panels a-c) by the negative clump migrating toward the rightward positive clump, and flux cancels at their mutual neutral line. Flux cancelation also goes on at that neutral line between  weaker clumps and grains that are located in-between the two larger canceling flux clumps (not shown here but visible in observations, e.g. in jet J6). (a) Flux cancelation via reconnection low above the photosphere  (red star in a) builds a highly sheared field (the sigmoid-shaped loop in b) and a short loop (brown loop in b). this low-altitude reconnection produces EUV brightening low along the neutral line. (b) The short-loop product of the reconnection (brown field line) submerges \citep{Balle89}. The pink star shows an interchange reconnection (that we call a `cool reconnection'; see text) in the low corona above the flux-cancelation neutral line. (c) The blue sigmoid-shaped curve represents a cool-plasma minifilament, which resides in the highly sheared field between the converging flux clumps. The cool reconnection resulted in the new open field line and the ew outer loop of the magnetic arcade over the filament neutral line; the newly-reconnected field lines are shown in red. The continued low-altitude flux cancelation eventually results in a jet-producing minifilament eruption; for the progression of the jet-producing eruption, see Figure 4 of \cite{panesar16b}.} \label{fig4}
\end{figure*}

Figure \ref{fig4} shows  a schematic interpretation of minifilament formation on the basis of our AIA and HMI observations. Each panel shows three flux clumps. Two of these, a majority-polarity (positive) flux clump  and a minority-polarity (negative) flux clump (the two right-most flux clumps) converge by the negative clump approaching the positive clump, which leads to photospheric-flow-driven flux-cancelation magnetic reconnection between them (shown by a red star in Figure \ref{fig4}a). We propose that this reconnection, which occurs low in the atmosphere above the neutral line, makes some brightenings at the neutral line before the minifilament formation; this would correspond to the pre-minifilament-formation brightenings we observe in Figures \ref{fig1b}a, \ref{fig2b}a, and  \ref{fig3b}a. This low-altitude flux-cancelation reconnection results in two reconnection products: a highly sheared field (the sigmoid-shaped loop in (b)), and a short loop (brown loop in (b)) across the neutral line. The short loop  submerges below the photosphere (indicated by the brown arrow) following \cite{Balle89}; this subduction would appear as the `cancelation' that we observe in the magnetogram movies. The cool minifilament plasma (blue sigmoid in c) collects in the highly sheared field (sigmoid-shaped reconnection product) above the neutral line of the two canceling flux clumps. In (b) and (c), as the middle flux clump moves rightward, it moves away from the majority-polarity (positive) flux clump on the left. This results in what we call here a  `cool reconnection' between the (left-most) highest closed lines and the (right-most) far-reaching field lines. (We do not observe any hot loops above the minifilament; that is why we call it cool reconnection (pink star in b); it starts before the existence of the minifilament, and does not produce any brightening or jet.) This cool reconnection builds two new fields: an overlying magnetic arcade enveloping the minifilament neutral line (red closed loop in (c)) and new open or far-reaching coronal field line (red curved line c). Meanwhile, the low photospherically-driven reconnection (red star in a) continues in the  flux-cancelation process at the neutral line, continually  adding additional shear and twist to the minifilament field.

The further evolution is not shown in the schematic here, but that further evolution continues in the schematic and discussion of \cite{panesar16b}, Figure 4. Briefly: Eventually, the twisted (or at least highly sheared) minifilament field becomes unstable and erupts outward due to the continuous flux cancelation at the neutral line. This results in the JBP via internal reconnection among the legs of the erupting minifilament field, and that erupting field (enveloping and carrying the minifilament) runs into neighboring open or far-reaching coronal field. This results in `external reconnection' between these fields, forming a hot spire along the open/far-reaching field, and with the cool minifilament material escaping into the jet spire. This is all as described in \cite{sterling15} and \cite{panesar16b}. One of the new findings from our study here is that quiet Sun homologous jet eruptions can occur if enough flux cancelation continues to occur at the minifilament neutral line. Eventually, the neutral line disappears,  no more minifilaments are produced, and so the homologous jets cease.

In conclusion, our observations show  that continuous flux cancelation between a minority-polarity flux clump and a majority-polarity flux clump builds a highly-sheared minifilament field, leading to the formation of a minifilament and its jet-driving eruption. Continued flux cancelation sometimes results in homologous eruptions. Thus we have found that magnetic flux cancelation is the key agent responsible for building the field for solar pre-jet minifilaments before triggering it to erupt to make a jet. These results are consistent with the models for the formation of the field for of typical solar filaments \citep{Balle89,martens01}.

\acknowledgments
N.K.P's research was supported by an appointment to the NASA Postdoctoral Program at the  NASA MSFC, administered by Universities Space Research Association under contract with NASA. A.C.S and R.L.M were supported by funding from the Heliophysics Division
of NASA's Science Mission Directorate through the heliophysics Guest Investigators Program, and by the \Hinode\ Project. We are indebted to the \sdo/AIA and \sdo/HMI teams for providing the high resolution data. \sdo\ data are courtesy of the NASA/\sdo\ AIA and HMI science teams.% We thank the referee for constructive comments.

\bibliographystyle{apj}

\end{document}